\documentclass[]{elsart}

\usepackage[numbers,sort&compress]{natbib}
\usepackage{amsmath}
\usepackage{bm}
\usepackage{nicefrac}

\begin{document}

\bibliographystyle{myprsty}

\begin{frontmatter}

\title{Instantons in Quantum Mechanics\\
and Resurgent Expansions}

\author{Ulrich D. Jentschura}
 
\address{\scriptsize Physikalisches Institut der Universit\"{a}t Freiburg,\\
Hermann--Herder--Stra\ss{}e 3, 79104 Freiburg im Breisgau, Germany and\\
National Institute of Standards and Technology, \\
Gaithersburg, MD20899-8401, Maryland, USA\\
{\bf electronic mail address: jentschura@physik.uni-freiburg.de.}}

\author{Jean Zinn--Justin}

\address{\scriptsize
DAPNIA/DSM
(D\'{e}partment d'astrophysique, de physique des particules,\\
de physique nucl\'{e}aire et de l'instrumentation associ\'{e}e),\\
Commissariat \`{a} l'\'{E}nergie Atomique, Centre de Saclay,
F-91191 Gif-sur-Yvette, France and\\
Institut de Math\'{e}matiques
de Jussieu--Chevaleret, Universit\'{e} de Paris VII, France\\
{\bf electronic mail address: zinn@spht.saclay.cea.fr.}}

\begin{abstract}
Certain quantum mechanical potentials
give rise to a vanishing perturbation series for at least
one energy level (which as we here assume
is the ground state), but the true ground-state energy
is positive. We show here that in a typical case,
the eigenvalue may be expressed in terms of a generalized
perturbative expansion (resurgent expansion). Modified
Bohr--Sommerfeld quantization conditions lead to generalized 
perturbative expansions which may be expressed in terms
of nonanalytic factors of the form $\exp(-a/g)$, where $a > 0$
is the instanton action, and power series in the coupling $g$,
as well as logarithmic factors. The ground-state energy, for 
the specific Hamiltonians, is shown to be dominated
by instanton effects, and we provide numerical evidence for the 
validity of the related conjectures.
\end{abstract}


\end{frontmatter}

\newpage

{\em Introduction.}---A number of intriguing and rather subtle issues are
connected with simple Rayleigh--Schr\"{o}dinger
perturbation theory when it is applied to certain 
classes of one-dimensional
quantum mechanical model problems, which give rise
to divergent perturbation series and allow for the 
presence of instanton effects~\cite{LGZJ1990}. 
Of particular interest is
the case of the symmetric double-well 
potential~\cite{BrPaZJ1977,Bo1980}
\begin{equation}
\label{Vgdw}
{\overline V}_{\rm dw}(g,q) =
{1 \over 2}\, q^2 \, (1 - \sqrt{g} \, q)^2 \,,
\end{equation}
the Hamiltonian being
$\overline{H}_{\rm dw} = -\nicefrac{1}{2}\, ({\rm d}/{\rm d}q)^2 +
{\overline V}_{\rm dw}(g,q)$.
There are several points to note: {\em (i)} The 
perturbation series can be shown to be non-Borel 
summable~\cite{BrPaZJ1977,Bo1980} for positive $g$. 
{\em (ii)} The parity operation
$q \to 1-q$ leaves ${\overline V}_{\rm dw}(g,q/\sqrt{g})$ invariant, 
and eigenfunctions are classified according to a principal
quantum number $N$ and the parity eigenvalue $\varepsilon=\pm$.
States with the same principal quantum number but opposite
parity are described by the same perturbative expansion.
{\em (iii)} The energy splitting between states of opposite parity 
is described by nonanalytic factors of the form $\exp[-1/(6 g)]$.
In general, quantum tunneling may 
generate additional contributions to eigenvalues of
order $\exp(-{\rm const.} \ /g)$, which have to be added to the
perturbative expansion (for a review and more detail about barrier
penetration in the semi-classical limit see for
example~\cite{ZJ1996ch43}).
Dominant contributions to the Euclidean path integral are generated
by classical configurations (trajectories) that describe quantum mechanical 
tunneling among the two degenerate minima; their Euclidean 
action remains finite in the limit of large positive
and large negative imaginary time (for a review see~\cite{Fo2000primer}).

Thus, the determination of eigenvalues starting from their
expansion for small $g$ is a non-trivial problem. 
Conjectures~\cite{ZJ1981jmp,ZJ1981npb,ZJ1983npb,ZJ1984jmp,Bo1994long} 
have been discussed in the literature
which give a systematic procedure to calculate eigenvalues, for
finite $g$, from expansions which are shown to contain powers of
the quantities $g$, $\ln g$ and $\exp(-{\rm const.}\ /g)$, 
i.e.~resurgent~\cite{Ec1981,St2002}
expansions. Moreover, generalized Bohr--Sommerfeld 
formulae (see e.g.~\cite[Eq.~(2)]{JeZJ2001}) can be
extracted by suitable transformations from the corresponding WKB
expansions. (The quantization conditions
may also be derived, approximatively, from an exact evaluation of
path integral in the limit of a vanishing instanton interaction,
by taking into account an arbitrary number of tunnelings between 
the minima of the potential~\cite{Bo1994,ZJJe2004i,ZJJe2004ii}.)
Note that the relation to the WKB expansion is not
completely trivial. Indeed, the perturbative expansion corresponds (from
the point of view of a semi-classical approximation) to a situation with
confluent singularities and thus, for example, the WKB expressions for
barrier penetration are not uniform when the energy goes to zero.

Here, we are concerned with a modification of the double-well
problem,
\begin{equation}                                                        
\label{VgFP}
{\overline V}_{\rm FP}(g,q) =
{1 \over 2}\, q^2 \, (1 - \sqrt{g} \, q)^2 + 
\sqrt{g}\,q - \frac12\,,
\end{equation}                                                        
the Hamiltonian being
$\overline{H}_{\rm FP} = -\nicefrac{1}{2}\, ({\rm d}/{\rm d}q)^2 +
{\overline V}_{\rm FP}(g,q)$.
The potential ${\overline V}_{\rm FP}(g,q)$
also contains a linear symmetry-breaking term.
There are the following points to note with regard to 
${\overline V}_{\rm FP}(g,q)$: {\em (i)} parity
is not conserved, and there is no degeneracy of the spectrum 
on the level of the perturbative expansion. {\em (ii)}~The perturbation
series for the ground state vanishes identically to all
orders in the coupling $g$~\cite{HeSi1978plb}. {\em (iii)} The true 
ground-state energy is positive; in~\cite{HeSi1978plb} it was
shown that it fulfills $0 < E_0 < C\,\exp(-D/g)$,
where $C$ and $D$ are positive constants.
Here, we present a resurgent expansion which 
naturally leads to a generalization of perturbation theory
valid for problematic potentials such as ${\overline V}_{\rm FP}(g,q)$.
Furthermore, we conjecture that a complete description
of the energy eigenvalues can be obtained via a generalized 
Bohr--Sommerfeld quantization condition which allows for 
the presence of nonanalytic contributions of order $\exp[-1/(3g)]$
for the ground state and of order $\exp[-1/(6g)]$ for excited states,
and we present numerical evidence for the validity of this 
conjecture. We thereby attempt to provide a complete 
description of the eigenvalues of the Fokker--Planck potential
by a generalized perturbation series involving instanton 
contributions. More general cases are treated in~\cite{ZJJe2004i,ZJJe2004ii}.

We are not concerned here with supersymmetric quantum mechanics.
In this context, the Fokker--Planck Hamiltonian has received some 
attention in the past two decades (see e.g.~\cite{Wi1981npb,AoEtAl1999}).
Instead, we rather attempt to find the suitable generalization of
perturbation theory that gives us an exact generalized secular
equation for the energy eigenvalues which in turn yields a 
generalization of perturbation theory suitable to the 
problem at hand. We will not satisfy ourselves
with an approximate solution of the problem but we attempt to find
complete expressions for the energy eigenvalues in terms
of resurgent expansions.

{\em Fokker--Planck Hamiltonian.}---The particularly interesting Hamiltonian 
${\overline V}_{\rm FP}(g,q)$ has been 
studied in \cite{HeSi1978plb}.
The spectra of the Hamiltonians $\overline{H}_{\rm dw}$ and 
$\overline{H}_{\rm FP}$ are invariant under the 
scale transformation $q \to q/\sqrt{g}$ and can 
therefore be written alternatively as
\begin{subequations}
\label{ehamdw}
\begin{eqnarray}
\label{econvention}
\label{hdw}
H_{\rm dw} &=& -{g\over2} \left(\frac{{\rm d}}{{\rm d}q}  \right)^2 +
{1\over g} \, V_{\rm dw}(q),
\\[1ex]
\label{vdw}
V_{\rm dw}(q) &=& {1 \over 2}\, q^2 \, (1 - q)^2\,,
\\[2ex]
\label{hfp}
H_{\rm FP} &=& -{g\over2} \left(\frac{{\rm d}}{{\rm d}q}  \right)^2 +
{1\over g} \, V_{\rm FP}(q),
\\[1ex]
\label{vfp}
V_{\rm FP}(q) &=& V_{\rm dw}(q) + g \left(q - \frac12\right)\,.
\end{eqnarray}
\end{subequations}
This is a representation which illustrates that
$g$ takes the formal role of $\hbar$ and that
the linear symmetry-breaking term in $V_{\rm FP}(q)$
in fact represents an explicit correction to the 
potential of relative order $g$.

For the double-well potential, the following two functions
enter into the generalized Bohr--Sommerfeld quantization 
formula~\cite{ZJ1984jmp,Bo1994,JeZJ2001},
\begin{subequations}
\begin{eqnarray}
B_{\rm dw}(E,g) &=& E+
g \left(3 E^2 + {\displaystyle{1\over4}}\right)
 + g^2 \left(35 E^3+{\displaystyle{25 \over4}}E\right) +
{\mathcal O}\left(g^3 \right),
\label{eBdble} \\
A_{\rm dw}(E,g) &=& {\displaystyle{1 \over 3g}}+
g \left(17E^2+{\displaystyle{19 \over 12}} \right)
+ g^2 \left(227\,E^3+{\displaystyle{187\over4}}E\right)+
{\mathcal O}\left(g^3 \right).
\label{eAdble}
\end{eqnarray}
\end{subequations}
The quantization condition and the resurgent expansion for
the eigenvalues read
\begin{equation}
\label{equantization}
{1 \over \sqrt{2\pi}} \, \Gamma\left( {1 \over 2} - B_{\rm dw}(E,g) \right) \,
\left(- {2 \over g} \right)^{B_{\rm dw}(E,g)} \,
\exp\left[-{A_{\rm dw}(E,g) \over 2}\right] = \varepsilon {\rm i}\,,
\end{equation}
and
\begin{eqnarray}
\label{ecomexp}
E_{\varepsilon,N}(g) &=&
\sum^{\infty}_{l = 0} E^{(0)}_{N,l} \, g^{l} 
\nonumber\\[2ex]
& & + \sum^{\infty}_{n=1} \left(2\over g\right)^{Nn}\,
\left( - \varepsilon {{\rm e}^{-1/6g}\over\sqrt{\pi g}} \right)^{n} \,
\sum^{n-1}_{k=0} \left\{ \ln\left(-\frac{2}{g}\right) \right\}^k \,
\sum^{\infty}_{l=0} e_{N,nkl} \, g^{l}.
\end{eqnarray}
Here, the $E^{(0)}_{N,l}$ are perturbative 
coefficients~\cite{ZJ1981jmp}, and the 
expression for $E_{\varepsilon,N}(g)$ follows naturally from 
an expansion of (\ref{equantization}) in powers of $g$,
$\ln(g)$, and $\exp[-1/(6g)]$.
The index $n$ characterizes the order of the instanton 
contribution ($n=1$ is a one-instanton, etc.).
The conjecture (\ref{equantization}) has been verified numerically
to high accuracy~\cite{JeZJ2001}.

Insight can be gained into the problem by considering the 
logarithmic derivative 
$S(q)=-g \, \psi'(q)/\psi(q)$, which for a general potential
$V$ satisfies the Riccati equation
\begin{equation}
g\,S'(q)-S^2(q)+2V(q)-2gE=0\,.
\label{eRiccat}
\end{equation}
This equation formally allows for solution with $E=0$
(and implies a vanishing perturbation series),
if the potential $V(q)$ has the following structure:
\begin{equation}
V(q) = \frac12\, \left[ U^2(q) - g \, U'(q) \right]\,.
\end{equation}
Indeed, a formal solution of $H \phi = 0$ in this case is given by
\begin{equation}
\label{ephi}
\phi (q) = \exp\left[ -{1\over g}\int^q {\rm d} q'U(q') \right] \,.
\end{equation}
The Hamiltonian $V_{\rm FP}$ is of this structure,
with $U(q) = U_{\rm FP}(q) = q\,(1-q)$.
This fact leads to the peculiar properties of $V_{\rm FP}$,
and indeed the Hamiltonians discussed in~\cite{HeSi1978plb}
belong to this class. The intriguing questions
raised by the remarks made in~\cite{HeSi1978plb} find a natural explanation
in terms of generalized Bohr--Sommerfeld quantization
conditions, and resurgent expansions.

Before discussing $V_{\rm FP}$, we first 
make a slight detour and consider the 
special case $U_{\rm II}(q) = q^3 + q$.
The potential $\textstyle{1 \over 2}\,U_{\rm II}^2(q)$ has no 
degenerate minima,
and thus there are no instantons to consider.
Indeed, in the case of the 
Hamiltonian $H_{\rm II} = -(g/2)\, ({\rm d}/{\rm d}q)^2 +
[U_{\rm II}^2(q) - g\,U'_{\rm II}(q)]/(2 g)$ (we follow the notation
of~\cite{HeSi1978plb}), the expression 
(\ref{ephi}) may be utilized for the 
construction of a normalizable eigenfunction of the Hamiltonian 
which reads $\phi_{\rm II}(q) = \exp[-(q^2/2 + q^4/4)/g]$
and has the eigenvalue $E=0$.

In the case of the 
potential $U_{\rm FP}(q) = q \,(1-q)$,
the issue is more complicated because the wave function
\begin{equation}
\label{epseudo}
\phi (q) = 
\exp\left[ \frac{1}{g} \, \left( \frac{q^3}{3} - \frac{q^2}{2} \right)
\right]
\end{equation}
is not normalizable, and thus is not an eigenfunction.
An analogy of the Riccati equation (\ref{eRiccat}) 
with the Fokker--Planck equation
suggests that the case $E=0$ be identified with an equilibrium
probability distribution.
Therefore, the non-normalizable wave function (\ref{epseudo})
may naturally be identified with
a ``pseudo-equilibrium'' distribution.

{\em Instanton action.}---The Euclidean instanton 
action for the ground state of the Fokker--Planck potential
is given by~\cite{ZJ1983npb,ZJ1984jmp} 
\begin{equation}
\label{edefA}
a = 2 \, \int_0^1 {\rm d} q\,U_{\rm FP}(q) = 2 \times \frac16 = \frac13\,,
\end{equation}
and it is this quantity which determines the 
leading contribution to the ground-state energy 
of order $\exp[-1/(3g)]$.
We conjecture here the following 
generalized quantization condition for the 
eigenvalues of the Hamiltonian (\ref{hfp})
\begin{equation}
\label{cfp}
{1\over \Gamma \left( - B_{\rm FP}(E, g)\right)\,
\Gamma \left(1 - B_{\rm FP}(E, g) \right)}
+ \left(-{2\over g}\right)^{2 B_{\rm FP}(E, g)} \,
\frac{\exp\left(- A_{\rm FP}(E, g)\right)}{2 \pi} = 0 \,.
\end{equation}
This condition is different from what would 
be obtained if one were to consider perturbation
theory alone. Indeed, the perturbative quantization condition reads
\begin{equation}
\label{econd}
B_{\rm FP}(E, g) = N\,,
\end{equation}
with integer $N \geq 0$. 
The functions $B_{\rm FP}$ and $A_{\rm FP}$ determine the perturbative
expansion, and the perturbative expansion about the instantons, in higher
order. They
have the following expansions,
%
%\begin{scriptsize}
\begin{subequations}
\begin{eqnarray}
\label{ebfp}
\lefteqn{B_{\rm FP}\left(E, g\right) = E +
3 \, E^2 \, g + \left( 35\, E^3 + \frac52\,E \right)\, g^2 }
\nonumber\\
& & + \left( \frac{1155}{2}\, E^4 + 105\,E^2  \right)\, g^3
+ \left( \frac{45045}{4}\, E^5 + \frac{15015}{4}\,E^3 
+ \frac{1155}{8}\,E \right)\, g^4
\nonumber\\
& & 
+ \left( \frac{969969}{4}\, E^6
+ \frac{255255}{2} \,E^4
+ \frac{111111}{8} \,E^2 \right)\, g^5
\nonumber\\
& & + \left( \frac{22309287}{4}\, E^7
+ \frac{33948915}{8}\,E^5 + \frac{3556553}{4}\,E^3
+ \frac{425425}{16}\,E \right)\,g^6\,,
\nonumber\\
& & + \left( \frac{2151252675}{16}\, E^8
+ \frac{557732175}{4}\,E^6 \right.
\nonumber\\
& & \qquad + \left. \frac{379257879}{8}\,E^4
+ 4157010 \,E^2 \right)\,g^7 +
{\mathcal O}(g^8)\,,
\\[2ex]
\label{eafp}
\lefteqn{A_{\rm FP}\left(E, g\right) = \frac{1}{3 g} +
\left( 17 \, E^2 + \frac56 \right)\, g
+ \left( 227 \, E^3 + \frac{55}{2} \, E \right)\, g^2 }
\nonumber\\
& &
+ \left( \frac{47431}{12} \, E^4
+ \frac{11485}{12}\,E^2
+ \frac{1105}{72} \right)\, g^3
\nonumber\\
& &
+ \left( \frac{317629}{4} \, E^5
+ \frac{64535}{2}\,E^3
+ \frac{4109}{2}\,E \right)\, g^4
\nonumber\\
& &
+ \left( \frac{26145967}{15} \, E^6
+ \frac{25643695}{24}\,E^4
+ \frac{4565723}{30}\,E^2
+ \frac{82825}{48} \right)\, g^5
\nonumber\\
& &
+ \left( \frac{812725953}{20}\, E^7
+ \frac{280162805}{8}\,E^5\right.
\nonumber\\
& & \left. \qquad + \frac{1057433447}{120}\,E^3
+ \frac{20613005}{48} \,E \right)\,g^6
+ {\mathcal O}(g^7)\,.
\end{eqnarray}
\end{subequations}
%\end{scriptsize}
%
The calculation of the functions $A$- and $B$-functions, for
general classes of potentials, is described in more detail 
in~\cite{ZJJe2004i,ZJJe2004ii}.
On the basis of (\ref{econd}) and (\ref{ebfp}),
we obtain the following perturbative expansion 
$E^{(\rm pert)}_N(g)$ up to and including terms of order $g^3$,
for general $N$,
\begin{equation}
\label{epert}
E^{(\rm pert)}_N(g) \sim N - 3 \, N^2 \, g
- \left( 17\, N^3 + \frac52\,N \right)\, g^2 
- \left( \frac{375}{2}\, N^4 + 
\frac{165}{2}\,N^2  \right)\, g^3
+ {\mathcal O}(g^4)\,.
\end{equation}
Here, the upper index $(0)$ means that only the perturbative
expansion (in powers of $g$) is taken into account.
For the ground state ($N=0$), all the terms vanish,
whereas for excited states with $N=1,2,\dots$, the perturbation series
is manifestly nonvanishing.

The quantization condition (\ref{cfp}) is conjectured to 
be the secular equation whose solutions determine
the the energy eigenvalues of the Fokker--Planck potential
(\ref{vfp}). The Eqs.~(\ref{ebfp}) and~(\ref{eafp}) 
can be used to expand the ground-state energy eigenvalue up to sixth order
in the nonperturbative factor $\exp(-1/3 g)$, and up to 
seventh order in the coupling $g$. The general structure of the 
resurgent expansions determined by (\ref{cfp}) differs slightly
for the ground state in comparison to the excited states.
This will be shown below, with a special emphasis on the 
ground state.

{\em Resurgent expansion for the ground state.}---Based on (\ref{cfp}), 
we derive the following expansion 
for the ground-state energy ($N=0$) of the Fokker--Planck 
potential~(\ref{vfp}):
\begin{eqnarray}
\label{eground}
E^{(0)}_{\rm FP}(g) &=&
\sum^{\infty}_{n=1} \left( {{\rm e}^{-1/3g} \over 2 \pi} \right)^{n} \,
\sum^{n-1}_{k=0} \left\{ \ln\left(-\frac{2}{g}\right) \right\}^k \,
\sum^{\infty}_{l=0} f^{(0)}_{nkl} \, g^{l}.
\end{eqnarray}
For small coupling $g$, this expansion is strongly dominated 
by the one-instanton effect ($n = 1$).
An explicit calculation using
(\ref{ebfp}) and (\ref{eafp}) leads to 
the following expansion for the ground-state energy of the 
Hamiltonian $H_{\rm FP}$, which is valid up to 
terms of order $[\exp(-1/3g)]^2$,
\begin{eqnarray}
\label{e0inst1}
E^{(0)}_{\rm FP}(g) &\sim&  
\frac{
\exp\left(-\frac{1}{3g}\right)}{2 \pi} \,
\left( 1 - \frac{5}{6}\,g - \frac{155}{72}\,g^2 
- \frac{17315}{1296}\,g^3 - \frac{3924815}{31104}\,g^4 \right.
\nonumber\\[1ex]
& & \left.
- \frac{3924815}{31104}\,g^4 
- \frac{294332125}{186624}\,g^5 
- \frac{163968231175}{6718464}\,g^6 
\right.
\nonumber\\[1ex]
& & \left. 
- \frac{18124314587725}{40310784}\,g^7 
- \frac{18587546509880725}{1934917632}\, g^8 +
{\mathcal O}(g^9)\right) 
\nonumber\\[1ex]
& & + {\mathcal O}\left( [\exp(-1/3g)]^2 \right)\,.
\end{eqnarray}
Because the perturbation series (\ref{epert}) vanishes for $N=0$,
the resurgent expansion starts with the one-instanton effect.
Indeed, Eq.~(\ref{e0inst1})
is the one-instanton contribution to the energy, characterized
by a nonanalytic factor $\exp(-1/3g)$ which 
is multiplied by a (divergent, nonalternating) power series in $g$.
This nonalternating series in $g$ may be resummed 
by a generalized Borel method (the generalized Borel sum
finds a natural representation in the sense
of distributional Borel summability,
which is effectively a Borel sum 
in complex directions of the parameters,
see e.g.~\cite{HeSi1978,FrGrSi1985,CaGrMa1993,Ca2000,Je2000prd}).

In analogy to the double-well potential, the 
imaginary part which is generated by this 
procedure (the ``discontinuity'' of the distributional Borel sum
in the terminology of~\cite{CaGrMa1993})
is compensated by an explicit imaginary part that stems from the 
two-instanton effect. We supplement here the first few terms
of the two-instanton shift of the ground-state eigenvalue 
[terms with $n=2$ in Eq.~(\ref{eground})]: 
\begin{eqnarray}
\label{e0inst2}
\lefteqn{\frac{[\exp(-1/3g)]^2}{(2 \pi)^2} \,
\left( 2 \ln\left( -\frac{2}{g}\right) + 2 \gamma \right.}
\nonumber\\[2ex]
& & \left. + g \left( -\frac{10}{3}\, \ln\left( -\frac{2}{g}\right)
-\frac{10}{3}\, \gamma - 3\right)
+ {\mathcal O}(g^2 \,\ln g) \right)\,.
\end{eqnarray}
Here, $\gamma = 0.577216\dots$ is Euler's constant.

The perturbative coefficients about one instanton,
called $f^{(0)}_{10K}$ in Eq.~(\ref{eground}), grow factorially
as 
\begin{equation}
\label{factorial}
f^{(0)}_{10K} \sim - \frac{3^K \, \Gamma(K)}{\pi} \,,
\qquad K \to \infty\,.
\end{equation}
It is an easy exercise to verify that this factorial
growth exactly leads to an imaginary part that is 
canceled by the imaginary part that results from 
the analytic continuation of the expression
$2 \ln(-2/g) + 2 \gamma$ in (\ref{e0inst2}) 
from negative to positive $g$. The explicit  coefficients 
in (\ref{e0inst1}) are consistent with the asymptotic 
formula (\ref{factorial}).

We have performed extensive numerical checks on the 
validity of the expansion (\ref{e0inst1}). For example, 
at $g = 0.007$, the ground-state energy, obtained numerically,
is
\begin{equation}
\label{numerics}
E^{(0)}_{\rm FP}(0.007) = 3.300\,209\,301\,936(1)\times 10^{-22}
\end{equation}
based on a calculation with a basis set composed of 
up to 300 harmonic
oscillator eigenstates. The numerical uncertainty is estimated
on the basis of the apparent convergence of the results under an 
appropriate increase of the number of states in the basis set.

When adding all terms up to the order of $g^9$
in the perturbative expansion
about the leading instanton (the first eight terms are
given in Eq.~(\ref{e0inst1}), further terms are available for 
download~\cite{JeHome}), we obtain 
\begin{equation}
E^{(0)}_{\rm FP}(0.007) \approx 3.300\,209\,301\,942\times 10^{-22}\,.
\end{equation}
With the term of order $g^{10}$ included, we have
\begin{equation}
E^{(0)}_{\rm FP}(0.007) \approx 3.300\,209\,301\,936\times 10^{-22}
\end{equation}
in full agreement with (\ref{numerics}) to all decimals shown.

We should clarify why the $n$-instanton contribution
in the resurgent expansion (\ref{ecomexp})
for the double-well potential (\ref{vdw})
involves the $n$th power of the expression $\exp(-1/6g)$, while in the
case of the ground-state of the 
Fokker--Planck Hamiltonian it involves
the $n$th power of $\exp(-1/3g)$. One may answer this question
by observing that in a symmetric potential, instanton configurations
with an odd number of tunnelings between the minima
yield a nonvanishing contribution to the path integral,
and therefore, the ``one-instanton'' configuration
in the double-well is a trajectory that starts in one well
and ends in the other. The linear symmetry-breaking
term of the Fokker--Planck potential lifts this degeneracy;
the leading, ``one''-instanton shift of the ground state
is now a configuration
in which the particle returns to the well from which it started;
the instanton action is therefore twice as large and
the two-instanton contribution (the ``bounce''-configuration
in the case of the double-well potential) becomes the
one-instanton solution in the case of the ground-state
of the Fokker--Planck equation.

As a last remark, it is useful to observe that
although the correction term $g(q-\nicefrac{1}{2})$
in Eq.~(\ref{vfp}) vanishes in the limit $g\to0$,
one cannot recover the double-well quantization condition
(\ref{equantization}) from (\ref{cfp}) in this limit;
it is nonuniform.

{\em Resurgent expansion for excited states.}---The energy
of excited states ($N > 0$) is dominated, for small $g$, by the 
perturbative expansion (\ref{epert}) which is manifestly 
nonvanishing to all orders in $g$. Because the symmetry is 
broken only at order $g$ [see Eq.~(\ref{vfp})], and because 
the dominance of the perturbation series (expansion in $g$)
is fully restored for excited states, the  resurgent expansion 
induced by (\ref{cfp}) becomes very close to the analogous 
expansion for the states of the double-well potential (\ref{ecomexp}).
By direct expansion of (\ref{cfp}), taking advantage of the 
the functional form of the dependence of $A_{\rm FP}(g)$ and 
$B_{\rm FP}(g)$ on $g$, we obtain the resurgent expansion
\begin{equation}
\label{eexcited}
E^{(\varepsilon, N>0)}_{\rm FP}(g) =
E^{(\rm pert)}_{N}(g) + \sum^{\infty}_{n=1} 
\left[- \varepsilon\,\Xi_N(g)\right]^n
\sum^{n-1}_{k=0} \left\{ \ln\left(-\frac{2}{g}\right) \right\}^k \,
\sum^{\infty}_{l=0} f^{(N)}_{nkl} \, g^{l}.
\end{equation}
Here, $\varepsilon = \pm$ is a remnant of the parity which 
is broken by $V_{\rm FP}(g)$, but only at order $g$, 
$E^{(\rm pert)}_{N}(g)$ is the perturbative expansion given 
in (\ref{epert}), and $\Xi_N(g)$ is given by
\begin{equation}
\Xi_N(g) = \sqrt{\frac{2}{\pi}}\,
\frac{2^{N-1}\,\exp\left(-1/6 g\right)}{g^N \,\sqrt{N!\,(N-1)!}}\,.
\end{equation}
For completeness, we indicate here the first few terms for the 
resurgent expansion of the states with $N=1$, but opposite
(perturbatively broken) parity $\varepsilon = \pm$, to leading
order in the coupling up to the three-instanton term,
\begin{eqnarray}
E^{(\pm, N=1)}_{\rm FP}(g) &=& 1 + {\mathcal O}(g)
\mp \Xi_1(g) (1 + {\mathcal O}(g)) \nonumber\\[2ex]
& & + \left[\Xi_1(g)\right]^2 \, \left( \ln\left(-\frac{2}{g}\right) +
\gamma - \frac12 + {\mathcal O}(g\ln g)\right)
\nonumber\\[2ex]
& & \mp \left[\Xi_1(g)\right]^3 \, 
\left( \frac32\,\ln^2\left(-\frac{2}{g}\right) +
\left( -\frac32 + 3 \gamma\right) \,\ln\left(-\frac{2}{g}\right) 
\right.
\nonumber\\[2ex]
& & \qquad \left. 
+ \frac58 - \frac32\,\gamma + \frac32\,\gamma^2 + \frac{\pi^2}{12} 
+ {\mathcal O}(g\ln g)\right)\,.
\end{eqnarray}

{\em Conclusions.}---We have presented the quantization condition (\ref{cfp}) 
which, together with Eqs.~(\ref{ebfp}) and (\ref{eafp}),
determines the resurgent expansions for an arbitrary state 
(quantum number $N$) of the Fokker--Planck
potential (\ref{hfp}) up to seventh order in the 
coupling $g$, and up to and including the six-instanton order.
For general $N > 0$, the perturbation series
is nonvanishing [see Eq.~(\ref{epert})], 
and the instanton contributions, for small coupling,
yield tiny corrections
to the energy. However, for the ground state
with $N=0$, the perturbation series vanishes to all orders in
the coupling, and the resurgent expansion~(\ref{eground}) for the 
ground-state energy of the Fokker--Planck potential (\ref{hfp})
is dominated by the nonperturbative factor $\exp(-1/3g)$ that 
characterizes the one-instanton contribution to the 
ground-state energy. The nonperturbative factor $\exp(-1/3g)$ 
is multiplied by a factorially 
divergent series [see Eqs.~(\ref{e0inst1}) and~(\ref{factorial})]; 
this is the natural structure
of a resurgent expansion which holds also for the 
double-well potential [see Eq.~(\ref{ecomexp})]. 
The basic features of this intriguing phenomenon 
have been described in~\cite{HeSi1978plb}; they
find a natural and complete explanation
in terms of the resurgent expansions discussed here.

Concepts discussed in the current
Letter may easily be generalized to 
more general symmetric potentials with degenerate
minima, potentials with two equal minima but asymmetric wells,
and periodic-cosine potentials (some further examples
are discussed in~\cite{ZJJe2004i,ZJJe2004ii}).
There is a well-known analogy between a one-dimensional field
theory and one-dimensional quantum mechanics, the one-dimensional
field configurations being associated with the
classical trajectory of the particle. 
Indeed, the loop expansion in field theory corresponds
to the semi-classical expansion~\cite[chapter~6]{ItZu1980}.
Therefore, one might hope that suitable 
generalizations of the methods discussed here
could result in new conjectures for problems where our present
understanding is (even) more limited.

Resurgent expansions appear to be of wide applicability
in a number of cases where ordinary perturbation theory, 
even if augmented by resummation prescriptions,
fails to described physical observables such as energy 
eigenvalues even qualitatively. This has been 
demonstrated here using the Fokker--Planck potential
as an example.

{\bf Acknowledgments.}
The authors would like to acknowledge the Institute of Physics,
University of Heidelberg, for the stimulating atmosphere during
a visit in January 2004, on the occasion of which part of
this work was completed, and the Alexander--von--Humboldt Foundation
for support. Roland Rosenfelder and Peter Mohr are gratefully
acknowledged for helpgul conversations.
The stimulating atmosphere at the National Institute
of Standards and Technology has contributed to
the completion of this project.


\begin{thebibliography}{10}

\bibitem{LGZJ1990}
J.~C. LeGuillou and J. Zinn-Justin, {\em Large-Order Behaviour of Perturbation
  Theory} (North-Holland, Amsterdam, 1990).

\bibitem{BrPaZJ1977}
E. Br\'{e}zin, G. Parisi, and J. Zinn-Justin, Phys. Rev. D {\bf 16},  408
  (1977).

\bibitem{Bo1980}
E.~B. Bogomolny, Phys. Lett. B {\bf 91},  431  (1980).

\bibitem{ZJ1996ch43}
J. Zinn-Justin, {\em Quantum Field Theory and Critical Phenomena}, 4th ed.
  (Clarendon Press, Oxford, 2002), ch.~43.

\bibitem{Fo2000primer}
H. Forkel, A Primer on Instantons, e-print hep-ph/0009136.

\bibitem{ZJ1981jmp}
J. Zinn-Justin, J. Math. Phys. {\bf 22},  511  (1981).

\bibitem{ZJ1981npb}
J. Zinn-Justin, Nucl. Phys. B {\bf 192},  125  (1981).

\bibitem{ZJ1983npb}
J. Zinn-Justin, Nucl. Phys. B {\bf 218},  333  (1983).

\bibitem{ZJ1984jmp}
J. Zinn-Justin, J. Math. Phys. {\bf 25},  549  (1984).

\bibitem{Bo1994long}
Several results have also been reported in J.~Zinn-Justin, contribution to the
  Proceedings of the Franco-Japanese Colloquium {\it Analyse alg\'e\-brique des
  perturbations singu\-li\`eres}, Mar\-seille-Luminy, October 1991, L.~Boutet
  de Monvel ed., Collection Travaux en cours, 47, Hermann (Paris 1994).

\bibitem{Ec1981}
J. \'{E}calle, {\em Les Fonctions R\'{e}surgentes, Tomes I---III, Publications
  Math\'{e}matiques d'Orsay, France (1981--1985)}.

\bibitem{St2002}
M. Stingl, Field--Theory Amplitudes as Resurgent Functions, e-print
  hep-ph/0207049.

\bibitem{JeZJ2001}
U.~D. Jentschura and J. Zinn-Justin, J. Phys. A {\bf 34},  L253  (2001).

\bibitem{Bo1994}
L. $\mathrm{Boutet~de~Monvel}$ (Ed.), {\em M\'{e}thodes R\'{e}surgentes}
  (Hermann, Paris, 1994).

\bibitem{ZJJe2004i}
J. Zinn-Justin and U.~D. Jentschura, Multi--Instantons and Exact Results~I:
  Conjectures, WKB Expansions, and Instanton Interactions, Ann. Phys. (N.Y.),
  at press (2004).

\bibitem{ZJJe2004ii}
J. Zinn-Justin and U.~D. Jentschura, Multi--Instantons and Exact Results~II:
  Specific Cases, Higher-Order Effects, and Numerical Calculations, Ann. Phys.
  (N.Y.), at press (2004).

\bibitem{HeSi1978plb}
I.~W. Herbst and B. Simon, Phys. Lett. B {\bf 78},  304  (1978).

\bibitem{Wi1981npb}
E. Witten, Nucl. Phys. B {\bf 185},  513  (1981).

\bibitem{AoEtAl1999}
H. Aoyama, H. Kukuchi, I. Okuochi, M. Sato, and S. Wada, Nucl. Phys. B {\bf
  533},  644  (1999).

\bibitem{HeSi1978}
I.~W. Herbst and B. Simon, Phys. Rev. Lett. {\bf 41},  67  (1978).

\bibitem{FrGrSi1985}
V. Franceschini, V. Grecchi, and H.~J. Silverstone, Phys. Rev. A {\bf 32},
  1338  (1985).

\bibitem{CaGrMa1993}
E. Caliceti, V. Grecchi, and M. Maioli, Commun. Math. Phys. {\bf 157},  347
  (1993).

\bibitem{Ca2000}
E. Caliceti, J. Phys. A {\bf 33},  3753  (2000).

\bibitem{Je2000prd}
U.~D. Jentschura, Phys. Rev. D {\bf 62},  076001  (2000).

\bibitem{JeHome}
See the URL {\tt http://tqd1.physik.uni-freiburg.de/\~{}ulj}.

\bibitem{ItZu1980}
C. Itzykson and J.~B. Zuber, {\em Quantum Field Theory} (McGraw-Hill, New York,
  NY, 1980).

\end{thebibliography}
\end{document}